\documentclass{iaus}
\usepackage{graphicx}

\title[Be stars in SMC open clusters] %% give here short title %%
{The WFI Halpha spectroscopic survey of the Magellanic Clouds: Be stars in SMC open clusters}

\author[C. Martayan \& D. Baade \& J. Fabregat]   %% give here short author list %%
{Christophe Martayan$^{1,2}$
 \and Dietrich Baade$^3$
 \and Juan Fabregat$^4$
 }

\affiliation{$^1$Royal Observatory of Belgium, 3 avenue circulaire \\ 1180 Brussels, Belgium 
 \\ email: {\tt martayan@oma.be} \\[\affilskip]
$^2$GEPI, Observatoire de Paris, CNRS, Universit\'e Paris Diderot; 
5 place Jules Janssen \\ 92195 Meudon Cedex, France \\[\affilskip]
$^3$ESO - European Organisation for Astronomical Research in the Southern Hemisphere, 
Karl-Schwarzschild-Str. 2, \\ D-85748 Garching b. Muenchen, Germany \\email: {\tt dbaade@eso.org}\\[\affilskip]
$^4$Observatorio Astron\'omico de Valencia, edifici Instituts d'investigaci\'o, 
Poligon la Coma, \\ 46980 Paterna Valencia, Spain \\email: {\tt Juan.Fabregat@uv.es}
}

\pubyear{2008}
\volume{256}  %% insert here IAU Symposium No.
\pagerange{119--126}
\jname{The Magellanic System: Stars, Gas, and Galaxies}
\editors{Jacco Th. van Loon \& Joana M. Oliveira, eds.}
\begin{document}

\maketitle

\begin{abstract}
At low metallicity, B-type stars show lower loss of mass and, therefore, angular momentum
so that it is expected that there are more Be stars in the Magellanic Clouds than
in the Milky Way.
However, till now, searches for Be stars were only
performed in a very small number of open clusters in the Magellanic Clouds.
Using the ESO/WFI in its slitless spectroscopic mode, we performed a
Halpha survey of the Large and Small Magellanic Cloud. Eight million
low-resolution spectra centered on Halpha were obtained. For their automatic analysis,
we developed the ALBUM code.
Here, we present the observations, the method to exploit the data and first
results for 84 open clusters in the SMC.
In particular, cross-correlating our catalogs with
OGLE positional and photometric data, we classified more than 4000 stars and were able
to find the B and Be stars in them.  We show the evolution of the rates
of Be stars as functions of area density, metallicity, spectral type, and age.

\end{abstract}

\firstsection % if your document starts with a section,
\section{Introduction}

Emission line stars (ELS) range from young to evolved stars (TTauri, Herbig Ae/Be, WR, Planetary Nebulae, etc), 
from hot to cool stars (Classical Be star, Oe, Supergiant star, Mira Ceti, Flare stars, etc). 
Among the ELS, here, we focus on the classical Be stars. They are non-supergiant B type stars, which have displayed at
least once emission lines in their spectra, mainly in the Balmer series of the hydrogen.
The emission lines come from a circumstellar decretion disk formed by episodic matter ejection of the central star.
It appears that the Be phenomenon is related to fast rotation and probably additional properties such as non-radial pulsation
or magnetic fields.  For a comprehensive review of Be stars in the Milky Way we refer the reader to \cite[Porter \& Rivinius (2003)]{}.
It seems also that the low metallicity plays a role (\cite[Kudritzki et al. 1987]{}): at low metallicity, 
typical of the Small Magellanic Cloud (SMC), the stellar radiatively driven winds are less efficient than at high
metallicity (typical of the Milky Way [MW]), thus the mass loss is lower and the stars keep more angular momentum. As a
consequence, B-type stars rotate faster in the SMC/LMC than in the MW (\cite[Martayan et al. 2007]{}).

It is then expected that the metallicity has also an effect on the Be-phenomenon itself as reported by 
\cite[Maeder et al. (1999)]{} or \cite[Wisniewski \& Bjorkman (2006)]{}, while the evolutionary phase could also play a role
(\cite[Fabregat \& Torrej\'on 2000]{}).
In most cases, the study of open clusters was done by using photometric observations, in the MW. 
The work by \cite[McSwain \& Gies (2005)]{} is a typical example.
To test these issues and improve the comparisons between SMC and MW, spectroscopic observations 
of stars in open clusters have to be done. 
In the SMC, a survey of emission line objects was performed by \cite[Meyssonnier \& Azzopardi (1993)]{} using photographic plates.
However, they were not able to study the stars in open clusters but mainly in the field. 
This paper deals with our slitless spectroscopic survey of ELS and Oe/Be/Ae stars in SMC open clusters, 
while in the whole SMC 3 million spectra were obtained. 

\begin{figure}[h!]
% \vspace*{-2.0 cm}
\begin{center}
 \includegraphics[width=3.1in, angle=-90]{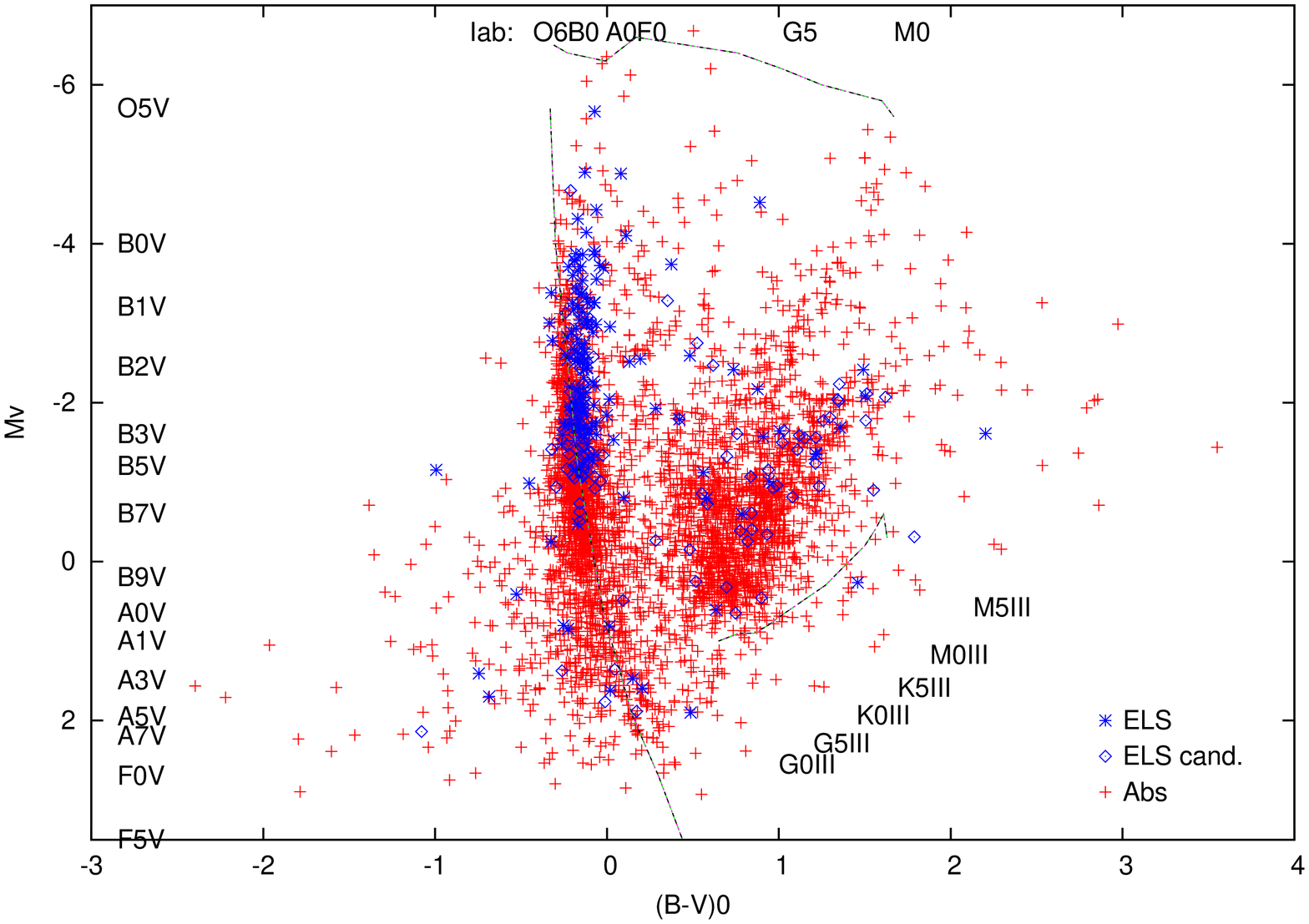} 
 \includegraphics[width=3.1in, angle=-90]{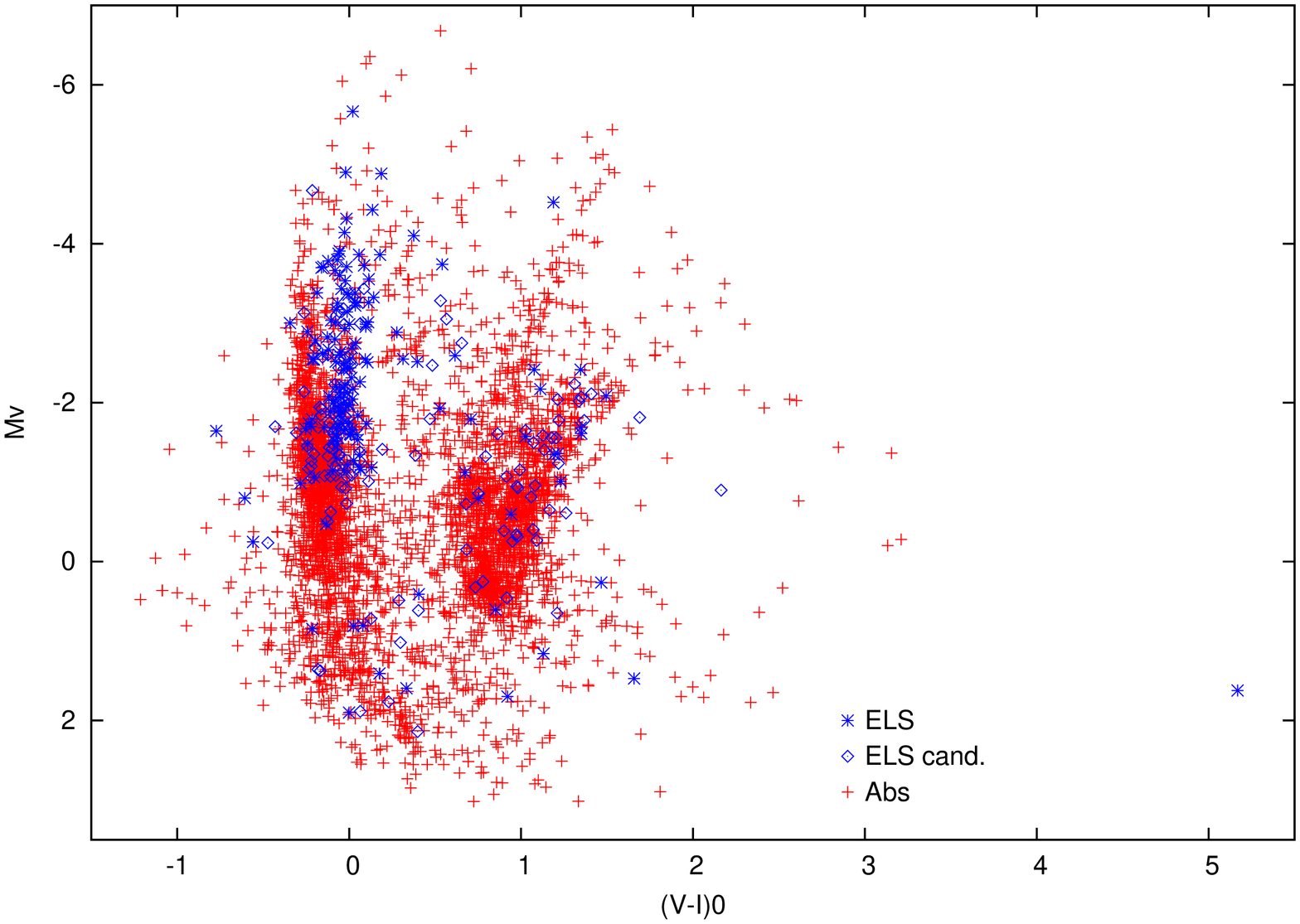} 
% \vspace*{-1.0 cm}
 \caption{Absolute V magnitude vs. dereddened colour (B-V)-top or (V-I)-bottom for SMC stars of our sample. 
 Blue crosses correspond to definite ELS, blue diamonds correspond to candidate ELS, and red + to absorption stars.}
   \label{figs1}
\end{center}
\end{figure}

%_________________________________________________________________________________________________________
\section{Observations, data-reduction}
To increase the number of open clusters studied in the SMC (1 to 6 in the previous photometric studies of 
\cite[Maeder et al. 1999]{} or \cite[Wisniewski \& Bjorkman 2006]{}),
to improve the statistics, and to quantify the evolution of the rates of Be stars to B stars with decreasing metallicity,
we performed a slitless spectroscopic survey of the SMC.  
The observations were obtained on September 25, 2002 with the ESO Wide Field Imager (WFI, see \cite[Baade et al. 1999]{}) at 
the 2.2m MPG Telescope located at La Silla in Chile.
We used its slitless spectroscopic mode with the R50 grism and a narrow filter centered on H$\alpha$ to reduce the crowding of 
observed areas.
We recall that this kind of instrumentation is not sensitive to the diffuse ambient nebulosities and does not allow
weak emission lines to be found.  With 10-minutes exposures with WFI, it is possible to detect emission lines with EW$\ge$10 \AA~or 
with a relative intensity to the continuum equal to 2 down to V magnitudes around 17.5.  For fainter stars, due to the noise in the spectra, 
it is only stronger emission lines are within reach. 
As an example of this kind of observations, see the study by \cite[Martayan et al. (2008a)]{}
in the MW NGC6611 open cluster and the Eagle Nebula. 

The CCD image treatment was performed using IRAF tasks. The spectra extraction was done using the SExtractor code
(\cite[Bertin \& Arnouts 1996]{}). The treatment of spectra and search for stars with emission were done by using
the ALBUM code (see \cite[Martayan et al. 2008b]{}).

Details about the samples used in this study are given in Table~\ref{tab1}. 

\begin{table}[h!]
  \begin{center}
  \caption{Details and comparison of samples used in this study}
  \label{tab1}
 {\scriptsize
  \begin{tabular}{|c|c|c|}\hline 
 & {\bf SMC (this study)} & {\bf MW from \cite[McSwain \& Gies (2005)]{}} \\ 
\hline
N open clusters & 84 & 54 \\
Definite Be stars & 109 & 52 \\
Candidate Be stars & 54 & 116 \\
Total calssical Be stars   & 163 & 168 \\
Ae stars & 7 & 57 \\
Oe stars & 6 & 3 \\
Other ELS (not MS) & 90 & \\
Unclassified ELS$^1$ & 49 & \\
NGC346$^2$ & 54 & \\
B stars & 1384 & 1741 \\
other stars & 2683 & 508 \\
\hline
  \end{tabular}
  }
 \end{center}
\vspace{1mm}
 \scriptsize{
 {\it Notes:}\\
  $^1$No OGLE photometry available for these stars. \\
  $^2$NGC346 is a complex young SMC open cluster, which contains both classical Be stars and Herbig Be/Ae stars but also other kinds
  of ELS (WR, TTauri).  Because of the risk of possible confusions, the classical Be stars from this cluster are not included in the statistics of Be stars. }
\end{table}

%__________________________________________________________________________________________________________
\section{Results}

After the data-reduction, the objects were sorted by categories (definite or candidate ELS, absorption stars), 
and the astrometry of the stars was performed with an accuracy of 0.5'' with the ASTROM package of \cite[Wallace \& Gray (2003)].\\

{\underline{\it Classification of stars}}.\\
We then cross-matched our catalogues with OGLE photometric catalogues in SMC open clusters from
\cite[Udalski et al. (1998)]{}, \cite[Pietrzynski \& Udalski (1999)]{}, and \cite[Udalski (2000)]{},  in order to obtain the 
B, V, I magnitudes of the stars, the reddening, and the age of open clusters. With this information, the absolute
magnitudes and the derreddened colour indices of the stars were derived.

Several open clusters are found with high ratios of Be to B stars, occasionally even higher than in NGC330, an open cluster
already known for its rich Be content.
In order to abstract from the large variations from one open cluster to another in their frequency of Be stars, the stars were grouped
in a global sample. We classified the 4300 stars in our sample by using the calibration provided by \cite[Lang (1992)]{}.
Fig.\,\ref{figs1} shows the corresponding global HR colour/magnitude diagrammes.
From these results it seems that the SMC sample is complete for the ELS till spectral types B3-B4, while for absorption stars it is
complete till B5.\\

\begin{figure}[h!]
% \vspace*{-2.0 cm}
\begin{center}
 \includegraphics[width=3.1in, angle=-90]{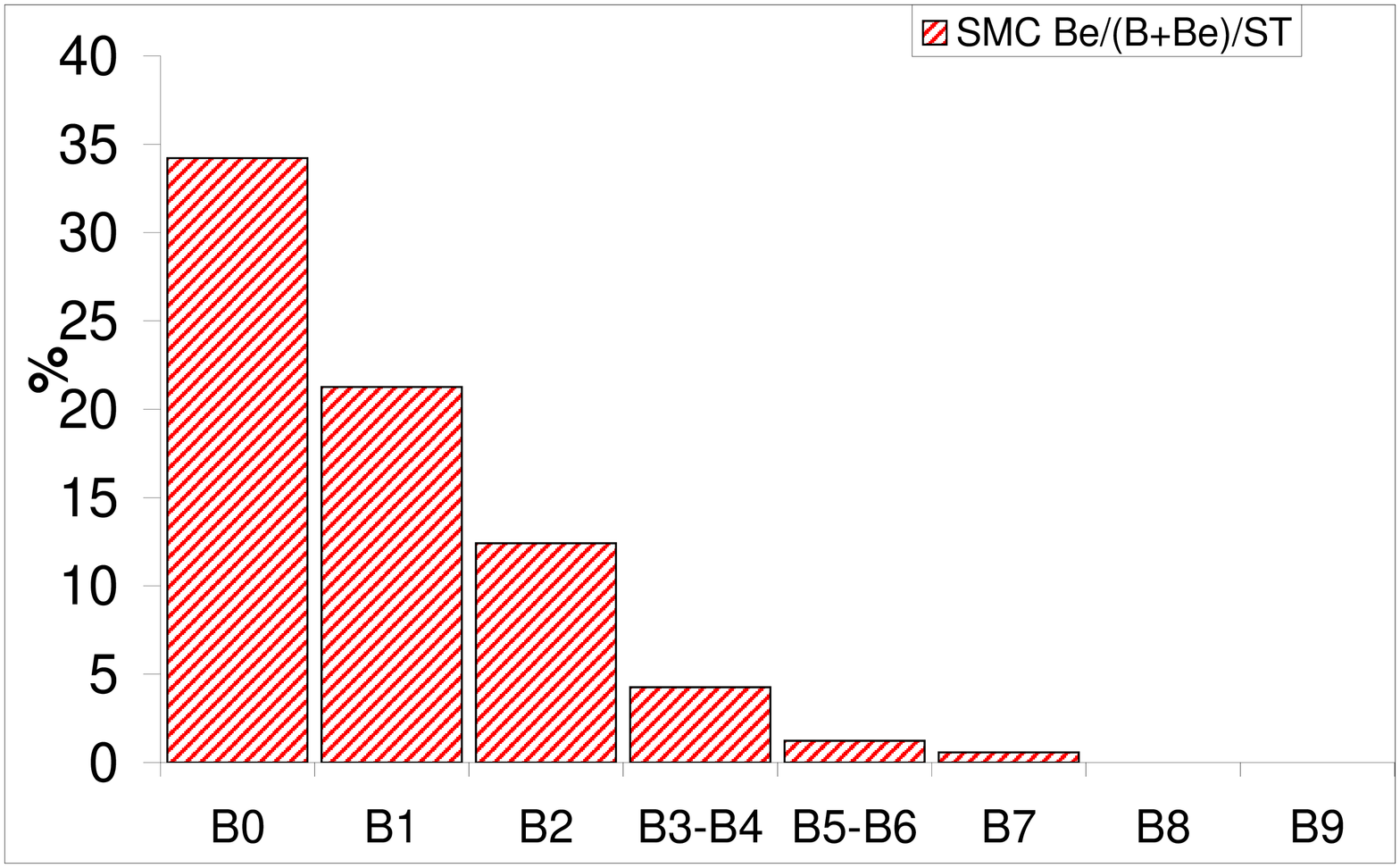} 
 \includegraphics[width=3.1in, angle=-90]{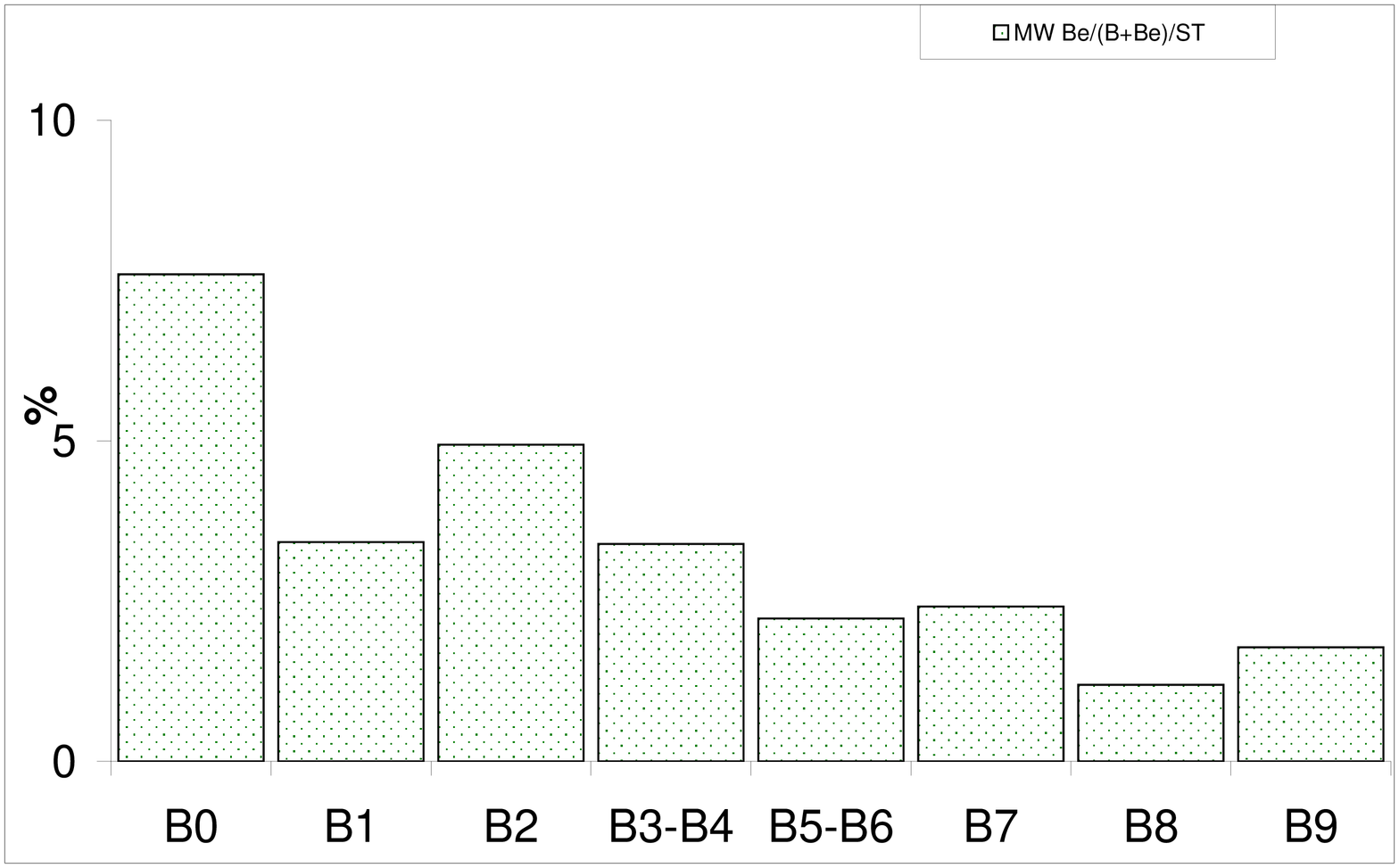} 
% \vspace*{-1.0 cm}
 \caption{Ratios of definite Be stars to B stars as a function of spectral type in the SMC (top), 
 and in the MW (bottom, from \cite[McSwain \& Gies 2005]{})}
   \label{figs3}
\end{center}
\end{figure}

%\begin{figure}[h!]
%% \vspace*{-2.0 cm}
%\begin{center}
% \includegraphics[width=3.4in, angle=-90]{fig3.ps} 
%% \vspace*{-1.0 cm}
% \caption{Number of Be stars in function of the density of open clusters in the SMC.}
%   \label{figs2}
%\end{center}
%\end{figure}
{\underline{\it Ratios of Be to B stars vs.\ metallicity}}.\\
To highlight and quantify a potential trend of the rates of Be stars to B stars with decreasing metallicity, we compared these
ratios by spectral-type categories with the study of \cite[McSwain \& Gies (2005)]{} in the MW. The same calibration 
for the classification of the stars was used in the MW. The comparison of the rates is shown in Fig.~\ref{figs3}.
Down to the completness limit of our study of ELS, one can see that the ratios are several times higher 
in the SMC than in the MW. Thus, there is an impact of the metallicity on the number of Be stars probably corresponding to the
increase of the rotational velocities in the SMC in comparison with the MW.\\

\begin{figure}[h!]
% \vspace*{-2.0 cm}
\begin{center}
 \includegraphics[width=3.1in, angle=-90]{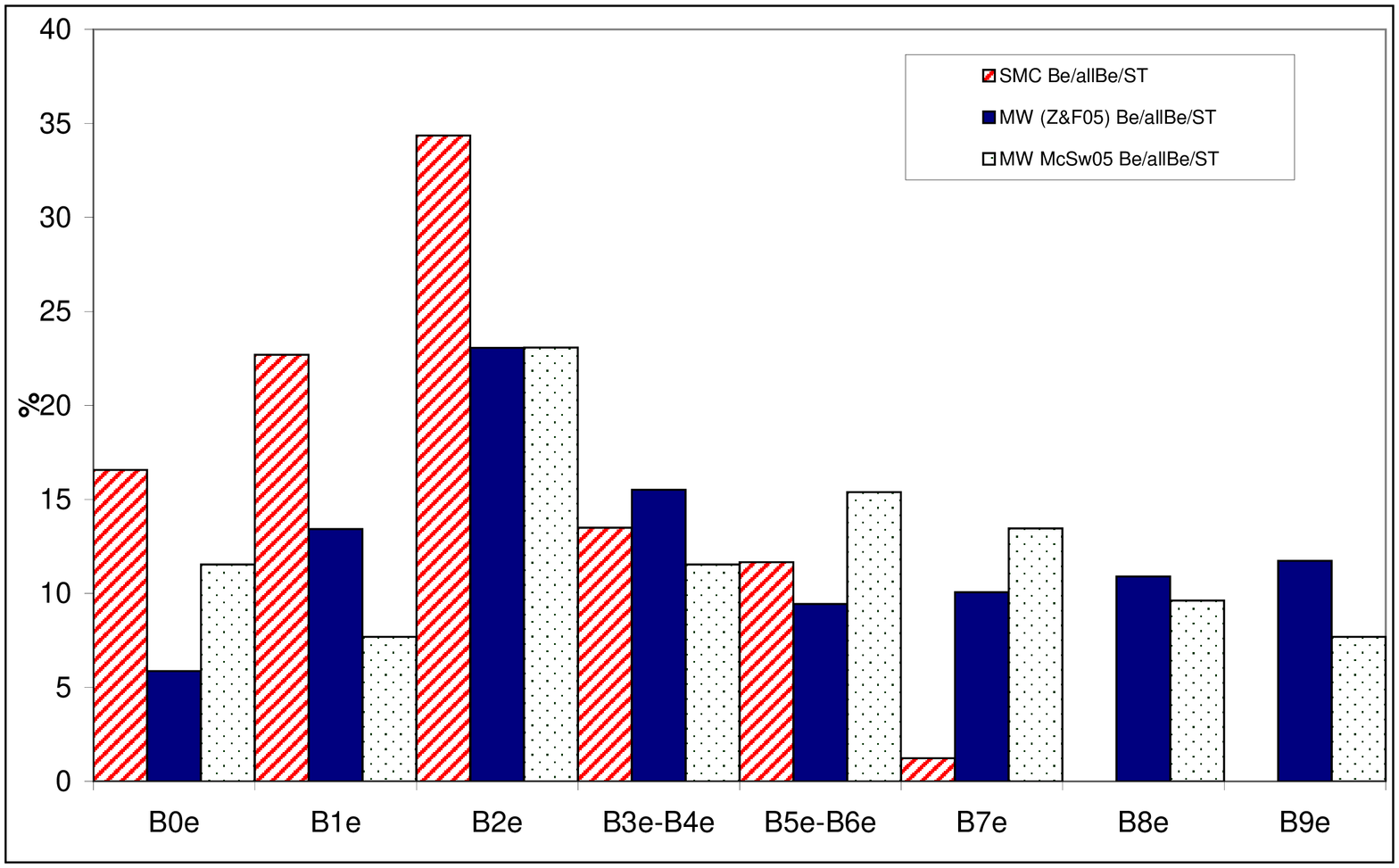} 
% \vspace*{-1.0 cm}
 \caption{Distribution of Be stars with spectral type in SMC (open clusters [hatched bars]) and  
 the MW (field [filled bars], and open clusters [dotted bars]).}
   \label{figs4}
\end{center}
\end{figure}

{\underline{\it Distribution of Be stars with spectral types}}.\\
The distribution of Be stars by spectral type categories in the SMC and MW was also studied.
In the MW, the results from \cite[McSwain \& Gies (2005)]{} in open clusters, and from 
\cite[Zorec \& Fr\'emat (2005)]{} in the field, were used.
The result is shown in Fig.~\ref{figs4}.
It appears that the maximum of Be stars is reached at the spectral type B2 in the SMC and MW.
There is also another peak at B5-B6, which can be seen in the MW studies (complete for the whole B-main sequence).
The first peak corresponds to the maximum of the emission intensity, 
the second one to the combined effects of the decrease of the emission intensity and the increase of the initial mass function of stars
towards late type stars.\\

{\underline{\it Distribution of open clusters with Be stars vs.\ age}}.\\
From preliminary results it appears that predominantly young open clusters\\
(age$<$100Myears) are found to contain classical Be (CBe) stars,
but depending on the types of the stars, certain CBe stars could have reached the terminal-age main sequence (TAMS). 
For example in the case of B0e stars, the TAMS is reached in $\sim$10 Myears. 
Some other old open clusters (age$>$100Myears) host CBe stars. From these results, it seems that certain
CBe could be born as CBe stars and some others appear during their evolution. 
This point needs more investigations star by star.

%________________________________________________________________________________________________________
\section{Conclusions}

Preliminary studies have shown a trend of the increase of the fraction of Be stars with decreasing metallicity.
However, up to now, the studies were only performed on a very limited sample of open clusters (less than 10 in the SMC) and  with
photometric data. Using the ESO/WFI in its slitless spectroscopic mode, we observed 84 open clusters in the SMC. 
Thanks to different codes and OGLE data, we were able to find and classify the emission line stars 
and absorption stars ($\sim$4300 stars).
The ratios of Be stars to B stars in the SMC and MW were studied. The comparison allows to quantify the increase 
of the number of Be stars with decreasing metallicity. Be stars in the SMC are 2 to 4 times more abundant than in the MW 
depending on the spectral types. It seems also that early Be stars
follow the same distribution in the SMC and MW with a maximum at the spectral type B2.
About the stellar phases at which Be stars appear, from our preliminary results, it seems that certain Be stars could be born 
as Be stars, while others could appear during the main sequence depending also probably on the metallicity and spectral types of stars.

\begin{acknowledgements}
C.M. acknowledges funding from the ESA/Belgian Federal Science Policy in the 
framework of the PRODEX program (C90290).
C.M. thanks support from ESO's DGDF 2006, 
the IAUS SOC/LOC for the IAUS grant, and the FNRS for the travel grant.
\end{acknowledgements}

\end{document}